\documentclass[10pt]{article}
\usepackage{mhchem}
\usepackage{fancyhdr}
\usepackage{extramarks} 
\usepackage{amsmath}
\usepackage{amsthm}
\usepackage{amsfonts}
\usepackage{siunitx}
\usepackage{tikz}
\usepackage[plain]{algorithm}
\usepackage{algpseudocode}
\usepackage{multirow}
\usepackage{booktabs}
\usepackage{palatino}
\usepackage{graphicx}
\usepackage{subfigure}
\usepackage[colorlinks,linkcolor=black,anchorcolor=black,citecolor=black,urlcolor=blue]{hyperref}
\usepackage{amsmath,bm}
\usepackage{booktabs}
\usepackage{mathtools}
\usepackage{amssymb}
\usepackage{caption}
\usepackage{capt-of}
\usepackage{mciteplus}
\usepackage{cite}
\usepackage{mathrsfs}
\usepackage{makecell}
\usepackage[title,titletoc,toc]{appendix}
\usepackage{xr}
\usepackage{parskip}
\usepackage{soul}
\usepackage{textcomp}
\usepackage[colaction]{multicol}
\usepackage[switch]{lineno}
\usepackage{lipsum}
\usepackage{etoolbox}
\usepackage{longtable}
\usepackage{array}
\usepackage{tablefootnote}
\usepackage{graphicx}
\usepackage{color, colortbl}
\usepackage{hhline}
\usepackage[T1]{fontenc} 

\definecolor{Lightblue}{rgb}{0.867,0.914,0.961}
\definecolor{Lightgreen}{rgb}{0.883,0.934,0.848}

\newcolumntype{C}[1]{>{\centering\arraybackslash}p{#1}}
\usetikzlibrary{automata,positioning}
\usetikzlibrary{automata,positioning}

\makeatletter
\newcommand*{\addFileDependency}[1]{% argument=file name and extension
	\typeout{(#1)}
	\@addtofilelist{#1}
	\IfFileExists{#1}{}{\typeout{No file #1.}}
}
\makeatother

\begin{document}
	\title{Multiscale differential geometry learning of networks with applications to single-cell RNA sequencing data}

\author{Hongsong Feng$^1$, Sean Cottrell$^1$, Yuta Hozumi$^1$, and  Guo-Wei Wei$^{1,2,3}$\footnote{
		Corresponding author.		Email: weig@msu.edu} \\% Author name
	\\
	$^1$ Department of Mathematics, \\
	Michigan State University, East Lansing, MI 48824, USA.\\
	$^2$ Department of Electrical and Computer Engineering,\\
	Michigan State University, East Lansing, MI 48824, USA. \\
	$^3$ Department of Biochemistry and Molecular Biology,\\
	Michigan State University, East Lansing, MI 48824, USA. \\
}
\date{\today} % Date for the report

\maketitle

	\textbf{\large Abstract:} Single-cell RNA sequencing (scRNA-seq) has emerged as a transformative technology, offering unparalleled insights into the intricate landscape of cellular diversity and gene expression dynamics. The analysis of scRNA-seq data poses challenges attributed to both sparsity and the extensive number of genes implicated. An increasing number of computational tools are devised for analyzing and interpreting scRNA-seq data. We present a multiscale differential geometry (MDG) strategy to exploit the geometric and biological properties inherent in scRNA-seq data. We assume that those intrinsic properties of cells lies on a family of low-dimensional manifolds embedded in the high-dimensional space of scRNA-seq data. Subsequently, we explore these properties via multiscale cell-cell interactive manifolds. Our multiscale curvature-based representation serves as a powerful approach to effectively encapsulate the complex relationships in the cell-cell network. We showcase the utility of our novel approach by demonstrating its effectiveness in classifying cell types. This innovative application of differential geometry in scRNA-seq analysis opens new avenues for understanding the intricacies of biological networks and holds great potential for network analysis in other fields.
	
	
	\pagenumbering{roman}
	\begin{verbatim}
	\end{verbatim}

	\newpage
	\clearpage
	\pagebreak
	%
	%{\setcounter{tocdepth}{4} \tableofcontents}
	\newpage
	
	\setcounter{page}{1}
	\renewcommand{\thepage}{{\arabic{page}}}

\section{Introduction}\label{sec:introduction}

Biological networks serve as abstract representations of biological systems, with biological entities or nodes representing genes, proteins, or metabolites, and edges signifying connections or relationships between corresponding biological entities. Critical networks in humans or animals encompass protein-protein interactions, protein-DNA interactions, protein-metabolite interactions, gene regulatory networks (GRNs), signal transduction networks, as well as metabolic and biochemical networks \cite{pavlopoulos2011using}. The complex biological network are intricately regulated by the dynamic expression levels of genes over both time and space. Consequently, the analysis of gene expression data plays a pivotal role in the realm of biological and medical research. For instance, it helps identify characteristic genes intimately linked to diverse cancer types and adeptly classifies tissue samples into distinct categories of normalcy and malignancy \cite{alharbi2023machine}.

 Single-cell RNA-sequencing (scRNA-seq) is a powerful and recent method for studying the gene expression of tens of thousands of single cells simultaneously, providing insights into the molecular states of individual cells through their transcriptional profiles \cite{lin2017using}. This technique represents a notable advancement over conventional bulk RNA-sequencing, which only evaluates average gene expression levels across a cell population. While average gene expression profiles are adequate for characterizing the overall tissue state, they mask signals from individual cells, thus overlooking tissue heterogeneity. The use of scRNA-seq technology has facilitated the discovery of new cell types \cite{usoskin2015unbiased}, the identification of novel markers for specific cell types \cite{usoskin2015unbiased, jaitin2014massively}, the exploration of cellular heterogeneity \cite{jaitin2014massively, pollen2014low}, and the trajectory inference of cellular differentiation \cite{luecken2019current}.

Despite its potential for revealing novel biological insights, scRNA-seq data present challenges such as sparsity, noisiness, and technical artifacts, which are beyond those encountered in bulk RNA samples \cite{poirion2018using}. Consequently, specific pre-processing and normalization methods tailored for scRNA-seq are essential. Commonly, scRNA-seq analysis involves dimension reduction techniques to mitigate noise and ensure computational tractability. However, the selection of dimension reduction method significantly impacts downstream analyses such as clustering \cite{zhu2017detecting} and pseudo-time reconstruction \cite{poirion2018using}. 

Principal component analysis (PCA) is a key technique in scRNA-seq and has been extensively used  for clustering single cells \cite{zhou2022pca}. PCA is a linear dimension reduction method that obtains a low-dimensional data representation along orthogonal linear axes, maximizing the proportion of variance accounted for on each axis in Euclidean space \cite{jolliffe2016principal}. Various methods have been designed based on PCA, such as pcaReduce, which uses a novel agglomerative clustering method atop PCA to cluster cells. Kernel PCA, employing kernel functions in a reproducing kernel Hilbert space, and accommodate non-linearity in data with complex algebraic/manifold structures. Some works combine graph Laplacians with PCA to incorporate nonlinear manifold in intrinsic geometrical  structure \cite{jiang2013graph} and generalize the traditional pairwise graph relations and capture multiscale geometrical structure in persistent Laplacian\cite{cottrell2023plpca}. In addition to PCA, other dimensionality reduction techniques for the analysis of scRNA-seq data include independent components analysis (ICA)\cite{trapnell2014dynamics}, Laplacian eigenmaps \cite{campbell2015laplacian}, t-distributed stochastic neighbor embedding(t-SNE) \cite{zeisel2015cell}, uniform manifoldapproximation and projection (UMAP) \cite{mcinnes2018umap}, deep learning methods \cite{lin2017using,luo2021topology}, and non-negative matrix factorization (NMF) methods \cite{shu2022robust,wu2020robust}.

Single-cell RNA sequencing (scRNA-seq) unveils gene associations and 
transcriptional networks within cell populations. Cell-cell and gene-gene 
networks from scRNA-seq aids understanding complex interactions and 
regulatory relationships. They are critical in cell type identification, the 
comprehension of tissue function through the analysis of ligand-receptor 
interactions inferred from scRNA-seq data \cite{armingol2021deciphering}, and 
deciphering co-expression patterns, the transcriptional regulatory landscape, 
and cell transitions states \cite{bourdakou2016discovering}. High-dimensional 
scRNA-seq data encompasses intricate manifolds. Mapper \cite{singh2007topological}  based on topological data analysis (TDA), offers a structural 
abstraction of often high-dimensional data and serves as a useful tool for 
scRNA data. Single-cell topological data analysis (scTDA)   was then developed to study temporal, unbiased transcriptional 
regulation using the scRNA seq data\cite{rizvi2017single}, and single-cell 
topological simplicial analysis (scTSA) was proposed based on temporal 
filtration to analyze single-cell genomics \cite{lin2022topological}. 
K-Nearest-Neighbors Topological PCA was proposed for processing  scRNA seq data \cite{cottrell2023k}. A topological NMF approach has been introduced for analyzing scRNA seq data \cite{hozumi2023analyzing}. These topological methods rely on algebraic topology, which involves the construction of a  simplicial complex and the computation of topological invariants. As a result, they effectively capture both the topological and 
geometric structures within scRNA-seq data. It is intriguing to preserve 
topological structures among cells in a low dimensional space and then 
interpret the biophysical meaning of high dimensional structure and geometry.

Differential geometry is a branch of mathematics that deals with the study of curves, surfaces, and other geometric objects using techniques from calculus and linear algebra. It provides a framework for understanding the intrinsic geometry of spaces and is widely used in various areas of mathematics and physics, including general relativity and differential topology. Differential geometry approaches have had  enormous success   in the multiscale modeling of  biomolecular systems  \cite{wei2010differential}, molecular surface representation  \cite{bates2008minimal}, and geometric learning  of  biomolecular properties \cite{nguyen2019dg}. Curvature analysis   offers useful quantitative measurement of low-dimensional manifold embedded in a high-dimensional ambient space and holds great promise for scRNA-seq data. Sritharan et al. \cite{sritharan2021computing} explored the 
viability of two approaches from differential geometry to estimate the 
Riemannian curvature of these low-dimensional manifolds. They applied their 
extrinsic approach to study scRNA-seq data of blood, gastrulation,and brain 
cells, and computed Riemannian curvature of scRNA-seq data manifold. Recent 
works \cite{zhou2021hyperbolic} investigated hyperbolic geometry to model cell-differentiation trajectories, which assumes the gene expression across 
multiple cell types exhibits a low-dimensional hyperbolic structure. Huynh et al.  
\cite{huynh2023topological} utilized Ricci curvature to analyze cell states 
in single-cell transcriptomic data on $k$-nearest neighborhood graphs built 
from PCA embeddings.  

In this study, we aim to introduce multiscale differential geometry (MDG) modeling of scRNA-seq data. We assume that the intrinsic biological and geometric properties of cells lies on a family of low-dimensional differentiable manifolds embedded in the high-dimensional scRNA-seq data. We employed our recently developed correlated clustering and projection (CCP)-assisted UMAP methodology to preprocess scRNA-seq data. This process involves conducting dimensionality reduction to project the data onto a low dimensional space, with the belief that the projection preserves faithful genetic variation, expression, and other biological information. To facilitate manifold analysis, we carry out a discrete-to-continuum mapping to obtain a volumetric density of the data,  rendering cell-cell interactive manifolds. This enables multiscale  differential geometry analysis by computing multiscale cell-cell interactive curvatures and elucidating cell–cell relationships in the original network. We demonstrated the efficacy of our multiscale differential geometry approach by classifying cell types using 12 scRNA-seq datasets, pairing the multiscale curvature representation of cells with a machine learning algorithm. Comparison results with predictive models based on embeddings from standard dimensionality reduction methods validate the descriptive and predictive ability of   multiscale differential geometry embeddings. Our multiscale differential geometry also possesses significant potential for offering insights to a variety of other networks in science and engineering.

\section{Method}\label{sec:method}

\begin{figure}[ht]
	\centering
	\includegraphics[width=0.85\linewidth]{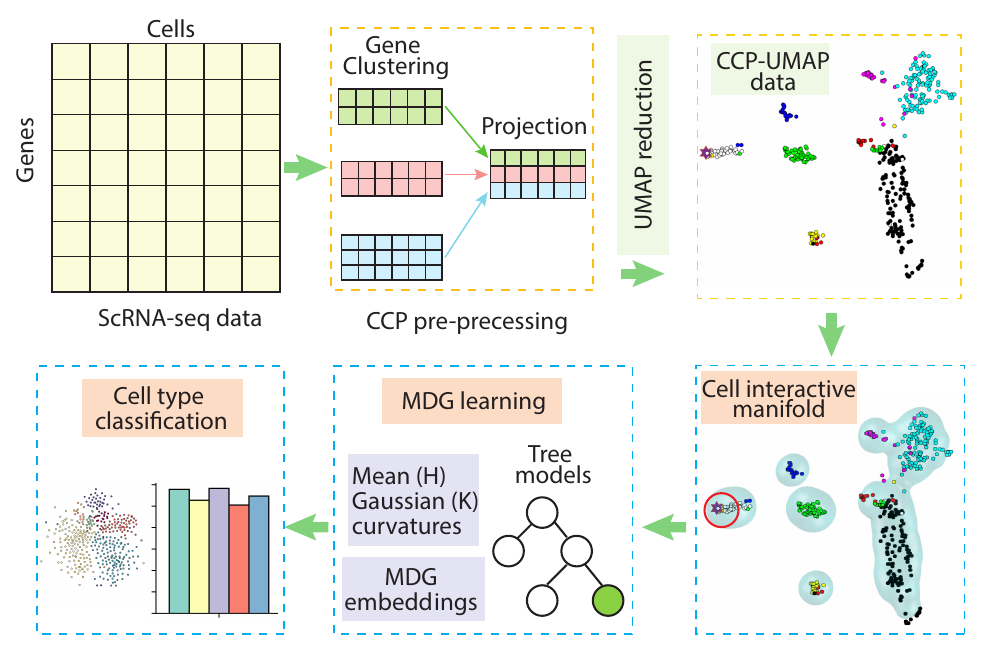} 
	\caption{{\footnotesize Illustration of the multiscale differential geometry (MDG) strategy for analyzing scRNA-seq data in the cell type classification task. Raw scRNA-seq data is preprocessed by correlated clustering and projection (CCP) approach, followed by UMAP dimensonality reduction. The resulting data forms the basis for constructing cell interactive manifolds, enabling the derivation of cell interactive curvatures through MDG analysis. The MDG modeling approach can then be employed for dissecting cellular heterogeneity, including tasks such as cell type classification.}}
	\label{Fig:concepts}
\end{figure}

\subsection{UMAP-assisted CCP Dimensionality reduction}

Dimensionality reduction is crucial for analyzing single-cell RNA sequencing (scRNA-seq) data. PCA and UMAP are two popular dimensionality reduction techniques and have been used in various applications including scRNA-seq data analysis \cite{tsuyuzaki2020benchmarking,becht2019dimensionality}. In particular, UMAP gained tremendous success and popularity in data visualization \cite{mcinnes2018umap}. Its most prominent strength lies in balancing both local and global structures of data. We use the dimensionality reduction methods for benchmarking analysis. We utilized Correlated Clustering and Projection (CCP) \cite{yuta2023singlecell}, a new data-domain dimensionality reduction method, to map our scRNA-seq data to lower dimensions. We further integrated the power of UMAP with CCP to interpret the original scRNA-seq data. Below we provide an overview of CCP and data abstraction procedures.	

The CCP procedure comprises two main steps: gene partitioning and gene projection. In essence, CCP maps clusters of similar genes into a collection of super genes that encapsulate pairwise nonlinear gene-gene correlations across all cells \cite{yuta2023singlecell}. This results in a super-gene representation. A preprocessing was performed first for the original scRNA-seq data. Assume $M$ to be the number of samples (cells) and $I$ to be the number of genes. Then we have the log-transformed scRNA-seq data $\mathcal{Z} \in \mathbb{R}^{M\times I}$. Each row represents expression information of all genes in each sample, while each column indicates the expression information of a gene in all samples.

\subsubsection{CCP Feature partitioning}

Let $\mathcal{Z} = \{\mathbf{z}^1, ..., \mathbf{z}^i, ..., \mathbf{z}^I\}$ be the collection of columns or gene vectors in $\mathcal{Z}$, and $\mathbf{z}^i \in \mathbb{R}^M$. CCP implements $k$-means clustering on the set $\mathcal{Z}$, resulting in clusters $Z^1, ..., Z^N$, $\mathcal{Z} = \uplus_{n=1}^N Z^n$, $N << I$. The clustering is performed on genes. Let $S = \{1, ..., I\}$ be the enumeration of the original genes. According to the $k$-means gene clustering results, we can have a gene number partition $S =\{S^1, ..., S^N\}$ with $S^n = \{i| \mathbf{z}^i \in Z^n\}$, i.e., $S^n$ is the set of gene numbers in the gene clustering $Z^n$. 

%Please change 	$\mathbb{R}^{|S^n|}$ to 	$\mathbb{R}^{S^n}$ in Section 2.1.2.

\subsubsection{CCP Feature projection}
With the gene partitioning, we establish the notation $\mathbf{z}_m^{S^n} \in \mathbb{R}^{|S^n|}$ to represent the set of $S^n$ genes in the $m$th sample. Here, $|S^n|$ indicate the cardinality of gene number set $S^n$. These genes are projected into a super-gene $x_m^n$ using the flexibility rigidity index (FRI) introduced in \cite{xia2013multiscale}. The key of FRI lies in the utilization of kernel functions. Denote $\|\mathbf{z}_i^{S^n} - \mathbf{z}_j^{S^n}\|$ as distance between  cell  $i$ and  cell $j$ for the cluster of  $S^n$ genes, and the gene-gene correlation between the two cells within the cluster of genes $S^n$ are defined by $\displaystyle C_{ij}^{S^n} = \Phi(\|\mathbf{z}_i^{S^n} - \mathbf{z}_j^{S^n}\|; \eta^{S^n}, \tau, \kappa)$, 
where $\Phi$ is a correlation kernel, and $\eta^{S^n}, \tau, $ and $ \kappa > 0 $ are parameters. Commonly employed metrics $||\cdot||$ for calculating the correlation include Euclidean, Manhattan, and Wasserstein distances.

The correlation kernels satisfy the following admissibility conditions
\begin{align}\label{eq:admissibility}
	& \Phi(\|\mathbf{z}_i^{S^n} - \mathbf{z}_j^{S^n}\|; \eta^{S^n}, \tau, \kappa) \to 0, \quad \text{as} \|\mathbf{z}_i^{S^n} - \mathbf{z}_j^{S^n}\| \to \infty \\ 
	& \Phi(\|\mathbf{z}_i^{S^n} - \mathbf{z}_j^{S^n}\|; \eta^{S^n}, \tau, \kappa) \to 1, \quad \text{as} \|\mathbf{z}_i^{S^n} - \mathbf{z}_j^{S^n}\| \to 0.
\end{align}
Commonly used kernel functions, such as radial basis functions, follow this pattern. Specifically, we utilize the generalized exponential function represented as:
\begin{align}\label{eq:exp-kernal}
	\Phi(\| \mathbf{z}\|; \eta) = e^{-(\| \mathbf{z} \| / \eta)^\kappa}, \kappa > 0, \eta >0,
\end{align}
where $\mathbf{z}$ is a general vector. 

We employ the following modeling approach for gene expression data:
\begin{align*}
	\Phi(\|\mathbf{z}_i^{S^n} - \mathbf{z}_j^{S^n}\|; \eta^{S^n}, \tau, \kappa) = \begin{cases}
		e^{-\left(\frac{\|\mathbf{z}_i^{S^n} - \mathbf{z}_j^{S^n}\|} { \eta^{S^n}\tau}\right)^\kappa} & \|\mathbf{z}_i^{S^n} - \mathbf{z}_j^{S^n}\| < r_c^{S^n} \\
		0, & \text{otherwise}.
	\end{cases}
\end{align*}
In this model, we introduce a cutoff distance denoted as $r_c^{S^n}$ and a scale parameter $\eta^{S^n}$, representing the resolution of gene-gene correlation and are determined by the data. The parameter $\kappa$ indicate the power of the exponential function, while $\tau$ serves as a scale parameter.

Pairwise gene-gene correlation matrix $C^{S^n} = \{C_{ij}^{S^n}\}$ reveals cell-cell similarity or interactions, and can also be mathematically perceived as the weighting of the edges in a weighted graph, given the cutoff $r_c^{S^n}$. The cutoff $r_c^{S^n}$ is determined as the 2-standard deviations of the pairwise distances.  On the other hand, $\eta^{S^n}$ can be regarded as the algebraic connectivity, defined as the average minimal distance between the cluster of genes 
\begin{align*}
	\eta^{S^n} = \frac{\sum_{m=1}^M \min_{\mathbf{z}_j^{S^n}}\|\mathbf{z}_m^{S^n} - \mathbf{z}_j^{S^n}\| }{M}.
\end{align*}
Leveraging the correlation function, we project $S^n$ genes into a super-gene using FRI for $i$th sample,
\begin{align*}
	x_i^n = \sum_{m=1}^M \Phi(\|\mathbf{z}_i^{S^n} - \mathbf{z}_m^{S^n}\|; \eta^{S^n}, \tau, \kappa).
\end{align*}
By performing this projection for all gene clusters, we get the lower dimensional super-gene representation for $i$th sample (cell), denoted as $\mathbf{x}_i = (x_i^1, ..., x_i^N)$.

The number $N$ in the super-gene representation $\mathbf{x}_i = (x_i^1, ..., x_i^N)$ indicates the gene clustering in the feature partitioning step, and It has impact on the representative ability of super-genes for each cell. In our previous study \cite{yuta2023singlecell}, we systematically investigated the performance of super-gene representation with different $N$ values. As $N$ increase, the super-gene representation can lead to better machine-learning predictions. When $N$ equals 300, the super-gene representations has proven to be effective and robust descriptors for the cells \cite{yuta2023singlecell}. 

\subsubsection{CCP-assisted UMAP dimensionality reduction}

UMAP is an essential dimensionality reduction technique. We further employ UMAP to carry dimension reduction on the aforementioned super-gene representations. The target dimension is 3D; specifically, the CCP super-gene representation $\mathbf{x}_i = (x_i^1, ..., x_i^N)$ is projected to a vector of length three. We still adopt notation of $\mathbf{x}$ for gene representation of cells. The resulting vector $\mathbf{x}_i = (x_i^1, x_i^2, x_i^3)$ is referred to as CCP-assisted UMAP (CCP-UMAP) representation. The set $\mathcal{X} = \{ \mathbf{x}_1, ..., \mathbf{x}_M\}$ becomes new gene representations for the $M$ cells, and $\mathcal{X}\in \mathbb{R}^{M,3}$. UMAP has the advantage of robustness when reducing data to lower dimensions, which underlies its popularity for data visualization.  In our recent investigations \cite{hozumi2023analyzing}, the integration of UMAP and CCP can significantly improve UMAP visualization and prediction accuracy in scRNA-seq data analysis. This motivates us to employ UMAP to project the CCP-preprocessed data onto a low dimensional gene representation. Subsequently, we utilize advanced mathematical tools for further analysis of scRNA-seq data. It is believed that CCP-UMAP representation $\mathcal{X}$ preserve sufficient information in CCP super-gene representations.

\subsection{Multiscale differential geometry modeling of cell-cell interactions}

Figure \ref{Fig:concepts} illustrates our  multiscale  differential geometry strategy for scRNA-seq data analysis. 
The CCP-UMAP representation provides a foundation for analyzing cell-cell similarity or interaction networks in low dimensional space. The resulting scRNA-seq data contain significant amount heterogeneity information in cells. Our objective is to utilize advanced differential geometry  modeling to capture interactions and correlations among cells, thereby revealing intrinsic heterogeneity information. To this end, we must convert discrete point cloud data into a density distribution by a discrete to continuum mapping. Then, manifolds are extracted and curvatures are evaluated.   

\subsubsection{Cell-cell interactive manifolds}

\begin{figure}[ht]
	\centering
	\includegraphics[width=0.75\linewidth]{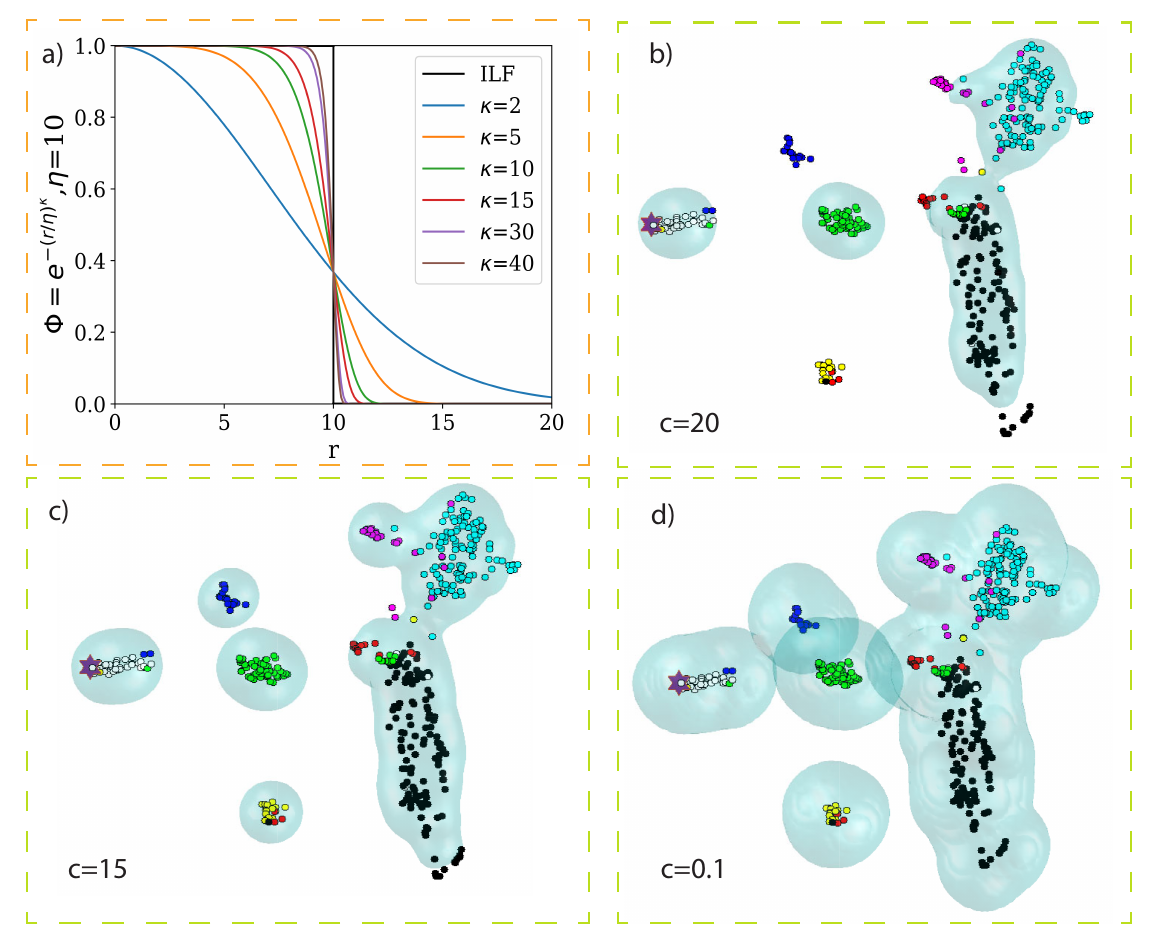} 
	\caption{{\footnotesize a. A demonstration of flexibility-rigidity correlation kernel defined by general exponential functions at different powers. At large $\kappa$ value, the kernel function essentially becomes an ideal low-pass filter (ILF). A larger $\eta$ value indicates a lower resolution and a slower decay. We can vary the values of $\eta$ and $\kappa$ to modulate the multiscaling modeling. b-d: The isosurfaces of a cell-cell interactive manifold for a cell (marked with hexagon ) in the scRNA-seq dataset GSE67835 at various isovalue values c. The parameters $\eta=1$ and $\kappa=8$ are used in the density function \eqref{eq:cell-cell correlation}.}}
	\label{Fig:exponential-filter}
\end{figure}

We first extract low-dimensional cell-cell interactive manifolds from the CCP-UMAP representations. This is done via a multiscale discrete to continuum mapping algorithm. We assume that single-cell properties are sampled on low-dimension manifolds defined by density functions. Given a cell-cell interactive manifold, we can apply tools from differential geometry to extract features suitable for machine learning tasks. 

In the CCP-UMAP representation $\mathcal{X} = \{ \mathbf{x}_1, ..., \mathbf{x}_M\}$, vector $\mathbf{x}_i$ represents $i$th cell, and $\| \mathbf{x} - \mathbf{x}_i \|$ is the Euclidean distance between a point $\mathbf{x} \in \mathbb{R}^3$ and the sample $\mathbf{x}_i$. The unnormalized gene expression density are given by a discrete to continuum mapping, 
\begin{equation}
	\rho(\mathbf{x}; \eta) = \sum_{i=1}^M \Phi(\| \mathbf{x} - \mathbf{x}_i \|; \eta),
\end{equation}
where $\Phi$ is a $C^2$ correlation kernel or density estimator. The correlation between a point with all cells are characterized by
\begin{equation} \label{eq:cell-cell correlation}
	\rho(\mathbf{x}; \eta)  = \sum_{i=1}^M e^{-(\| \mathbf{x} - \mathbf{x}_i \| / \eta)^\kappa}, \kappa > 0,
\end{equation} 
if $\Phi$ is defined to be the exponential kernel function $\eqref{eq:exp-kernal}$.

The kernel function reflects the topological connectivity of a cell-cell network or graph consisting of cells and characterizes the geometric compactness of the connectivity network. We construct a multiscale geometric representation based on the FRI theory, which cast the point cloud data into a density map. The conversion is achieved using resolution parameters $\eta$ and $\kappa$, which facilitates the multiscale analysis of complex data. As shown in Figure \ref{Fig:exponential-filter}a, the exponential kernal function show different correlation scales under various power values $\kappa$. It demonstrate the decay property of exponential and or other density functions. A larger $\eta$ value indicates a lower resolution and a slower decay. Note that other $C^2$ correlation kernels or density estimators can be utilized so long as they meet the admissibility conditions \eqref{eq:admissibility}.

The correlation function $\rho$ is regulated by scale parameter $\eta$ and decay parameter $\kappa$. By choosing multiple $\eta$ and $\kappa$ values, multi-scale characterizations are induced. In our multiscale modeling, we set a lower  and upper bound for $\eta$ as
\begin{align}\label{eq:etabounds}
	\eta_{\min} = \lceil \frac{\sum_{m=1}^M \min_{\mathbf{x}_i}\|\mathbf{x}_m - \mathbf{x}_i\| }{M}\rceil, \quad \eta_{\max} =\lceil \frac{\sum_{m=1}^M \max_{\mathbf{x}_i}\|\mathbf{x}_m - \mathbf{x}_i\| }{M}\rceil,
\end{align}
where notation $\lceil \cdot \rceil$ indicates the ceiling of the average distances, and $x_i\in \mathcal{X}$. This way, the correlation function \eqref{eq:cell-cell correlation} gives cell-cell interactive manifold representations. Figure \ref{Fig:concepts} offers a visualization of a cell-cell interaction manifold generated by specifying a particular isovalue of the kernel function \eqref{eq:cell-cell correlation} for a designated cell, indicated by a hexagon. The shape evolution of isosurfaces for the manifold through varying isovalues can be observed Figure \ref{Fig:exponential-filter}.

\subsubsection{Multiscale differential geometry of differential manifolds}

The aforementioned density functions renders the interactive manifold, which makes it feasible to use differential geometry to interpret cell-cell interactions. We have defined calculus on differentiable manifolds.

Consider a $C^2$ mapping $\mathbf{M}: U \rightarrow \mathbb{R}^{n+1}$, where $U \subset \mathbb{R}^n$ is open and the closure of $U$ is compact \cite{bates2008minimal}. Here, $\mathbf{M}(\mathbf{u}) = (M_1(\mathbf{u}), ..., M_n(\mathbf{u}), M_{n+1}(\mathbf{u}))$ is a position vector on a hypersurface, and $\mathbf{u} = (u_1,...,u_n) \in U$. Tangent vectors or directional vectors of $\mathbf{M}$ are $V_i = \frac{\partial \mathbf{M}}{\partial u_i},  i =1,...n$. The Jacobian matrix of $\mathbf{M}$ is $D\mathbf{M} = (V_1,V_2,...,V_n$). Given the notation $<\cdot>$ as the Euclidian inner product in $\mathbb{R}^{n+1}$, the first fundamental form $I$ is then given by $ I(V_i, V_j) :=<V_i,V_j>$ for any two tangent vector $V_i,V_j\in T_{\mathbf{u}}\mathbf{I}$, where $T_{\mathbf{u}}\mathbf{I}$ is the tangent hyperplane at $\mathbf{I(u)}$. In the coordinate $\mathbf{I(u)}$, the first fundamental form is a symmetric and positive matrix  $(g_{ij})= (I(V_i,V_j))$.

Let $\mathbf{N}(\mathbf{u})$ be the unit normal vector given by the Gauss map $N: U \rightarrow \mathbb{R}^{n+1}$
\begin{equation}
	N(u_1,...,u_n) = V_1 \times V_2 \times ... \times V_n / \|V_1 \times V_2 \times ... \times V_n \| \in \perp_\mathbf{u}\mathbf{M},
\end{equation}
where $\times$ denotes the cross product in $\mathbb{R}^{n+1}$ and $\perp_\mathbf{u}\mathbf{M}$ is the normal space of $\mathbf{M}$ at point $\mathbf{p}=\mathbf{M}(\mathbf{u})$. The normal vector $\mathbf{N}$ is orthogonal to the tangent hyper-plane $T_\mathbf{u}\mathbf{M}$ at $\mathbf{M}(\mathbf{u})$. With the normal vector $\mathbf{N}$ and the tangent vector $V_i$, we can define the second fundamental form: 
\begin{equation}
	II(V_i,V_j) = (h_{ij})_{i,j=1,2,\cdots,n} = (\langle - \frac{\partial N}{\partial u_i}, V_j\rangle )_{ij},
\end{equation}
The mean curvature can be calculated by $H = h_{ij}g^{ji}$, where we use the Einstein summation convention, and $g^{ji}=(g_{ij})^{-1}$. In addition, the Gaussian curvature is given by $K = \frac{\text{Det}(h_{ij})}{\text{Det}(g_{ij})}$.

\subsubsection{Multiscale cell-cell interactive curvatures}

With the above theory of differentiable manifold, both the mean curvature $H$ and Gaussian curvatures can be utilized to measure the cell-cell correlations based on the interactive manifolds $\rho$. The Gaussian curvature $K$ and mean curvature $H$ for the interactive manifold can be evaluated as below \cite{xia2014multiscale}:
\begin{align}\label{:eq:gaussain curvature}
	K =& \frac{2\rho_x\rho_y\rho_{xz}\rho_{yz}+2\rho_x\rho_z\rho_{xy}\rho_{yz}+2\rho_y\rho_z\rho_{xy}\rho_{xz}}{g^2}
	-\frac{2\rho_x\rho_z\rho_{xz}\rho_{yy}+2\rho_y\rho_z\rho_{xx}\rho_{yz}+2\rho_x\rho_y\rho_{xy}\rho_{zz}}{g^2}\\
	&+\frac{\rho_z^2\rho_{xx}\rho_{yy}+\rho_x^2\rho_{yy}\rho_{zz}+\rho_y^2\rho_{xx}\rho_{zz}}{g^2}-\frac{\rho_x^2\rho_{yz}^2+\rho_y^2\rho_{xz}^2+\rho_z^2\rho_{xy}^2}{g^2}\notag
\end{align}
where $g=\rho_x^2+\rho_y^2+\rho_z^2$. The mean curvature is the average second derivative with respect to the normal direction,
\begin{align}\label{:eq:mean curvature}
	H = \frac{1}{2g^{\frac{3}{2}}}[2\rho_x\rho_y\rho_{xy}+2\rho_x\rho_z\rho_{xz}+2\rho_y\rho_z\rho_{yz}
	-(\rho_y^2+\rho_z^2)\rho_{xx}-(\rho_x^2+\rho_z^2)\rho_{yy} -(\rho_x^2+\rho_y^2)\rho_{zz}\notag
	],
\end{align}
Meanwhile, the minimum curvature $\mu_{\text{min}}$, and maximum curvature, $\mu_{\text{max}}$,  can be calculated by: 
\begin{align*}
	\mu_{\text{min}} = H - \sqrt{H^2-K}\\
	\mu_{\text{max}} = H + \sqrt{H^2-K}
\end{align*}
In differential geometry, there are several important curvature concepts used to measure how much a geometric object deviates from being flat. These concepts can be applied to curves, surfaces, and higher-dimensional manifolds. In biomolecular studies, they have played important roles in stereo specificity of biomolecular surfaces \cite{cipriano2009multi}, protein–protein interaction hot spots, ligand binding pockets \cite{feng2012geometric}. Mean curvature and gaussian curvature can be suitable descriptions for cell-cell interactions.

Given the density function $\rho$, the Gaussian and mean curvatures are continuous and can be analytically determined. This renders their expressions free of numerical errors and directly suitable for cell-cell modeling. Furthermore, the associated computational expense is low due to the fast decaying effect of the kernel within a short manifold band \cite{opron2014fast}.

\subsubsection{Multiscale differential geometry modeling of cell-cell interactions}

Given the CCP-UMAP representation, the interactive manifold \eqref{eq:cell-cell correlation} enables us to utilize our differential geometry for cell-cell interaction analysis. The aforementioned Gaussian and mean curvatures provide a quantitative measure $(K,H)$ of a sample cell's interaction with other cells. 
By adjusting the values of $\eta$ and $\kappa$ in the correlation function \eqref{eq:cell-cell correlation}, we gather a collection of Gaussian and mean curvatures, thereby achieving multiscale modeling. The encoded mutliscale differential geometry information become suitable inputs for machine learning predictions.

Specifically, suppose we have a set of $\eta$ values, $\eta_i$ for $i=1,2,\cdots,p$, and a set of $\kappa$ values, $\kappa_j$ for $j=1,2,\cdots,q$. For the $m$th cell among the $M$ cells, there is a curvature vector $C_m=\{(K_m^{i,j},H_m^{i,j})|i=1,2,\cdots,p;j=1,2,\cdots,q\}$ and $C_m\in \mathbb{R}^{2p\cdot q}$. The curvature set $\mathcal{C}=\{C_1,C_2,\cdots,C_M\}$ serves as cell-cell interaction features and $\mathcal{C}\in \mathbb{R}^{M\times(2p\cdot q)}$. We call such multiscale differential geometry  modeling of cell-cell interactions or network analysis as MDG.

\subsection{Residue-similarity analysis}

Residue-Similarity (R-S) scores and indexes was proposed in our previous work \cite{yuta2022ccp}. It can be an alternative visualization approach in addition to available dimensionality reduction techniques. Assume the interested data with a set $\Omega=\{(\mathbf{x}_m,y_m)|\mathbf{x}_m \in \mathbb{R}^N, y_m \in \mathbb{Z}_L\}_{m=1}^M$, where $\mathbf{x}_m$ is the $m$th data point. The label $y_m$ indicates the ground truth or cluster label in classification or clustering problem. The dataset has $M$ data samples and $\mathbf{x}_m \in \mathbb{R}^N$  is the feature representation. $L$ indicate the number of data types, namely, $y_m \in [0,1,2,\cdots,L]$ . The whole set $\Omega$ is partitioned into $L$ classes 
$\omega_l=\{\mathbf{x}_m \in \Omega|y_m=l\}$ according to the labels $y_m=l$ and hence $\Omega= \cup_{i=0}^{L-1} w_l.$

There are two components in R-S scores. The residue score is defined to be the normalized inter-class sum of the distances. Suppose $\mathbf{x}_m \in \omega_l,$ the inter-class sum of the distances is defined to be
\begin{align*}
	R(\mathbf{x}_m) = \sum\limits_{\mathbf{x}_j\notin \omega_l} ||\mathbf{x}_m-\mathbf{x}_j||,
\end{align*}
where $||\cdot||$ is a certain distance metric. The residue score for $\mathbf{x}_m$ is given as
\begin{align}
	R_m := \frac{1}{R_{\max}}R(\mathbf{x}_m),
\end{align}
where $R_{\max}=\max \limits_{\mathbf{x}_m \in \Omega}R(\mathbf{x}_m).$ The similarity score $S_m$ is the average of the intra-class scores. Specifically, for any $\mathbf{x}_m \in \omega_l$,
\begin{align}
	S_m := \frac{1}{|\omega_l|} \sum\limits_{\mathbf{x}_j\in \omega_l} \left(1-\frac{||\mathbf{x}_m-\mathbf{x}_j||}{d_{\max}}\right),
\end{align}
where $d_{\max}= \max \limits_{\mathbf{x}_i,\mathbf{x}_j\in\Omega}||\mathbf{x}_i-\mathbf{x}_j||$. Both the residue score and similarity score range between 0 and 1. The Euclidean distance can be a metric to define the R-S scores. Generally, a large residue score $R_m$ indicates the data has large dissimilarity from other class data, while a high similarity score $S_m$ suggests that the data in the same class is well-clustered.

The class residue index (CRI) and class similarity index can be defined for each class. For a class $\omega_l$, the two indexes are defined as CRI$_l= \frac{1}{|w_l|}\sum_m R_m$ and CSI$_l= \frac{1}{|w_l|}\sum_m S_m.$ Their range are in $[0,1]$. Analogously, we have residue index (RI) and similarity index (SI) for the whole set as RI$=\frac{1}{L}\sum_l $CRI$_l$ and SI$=\frac{1}{L}\sum_l $CSI$_l$. 

Furthermore, with Residue Index (RI) and the Similarity Index (SI), we can calculate the disparity in residue similarity by simply taking $\text{RSD} = \text{RI} - \text{SI}$. It is denoted as Residue Similarity Disparity (RSD). Besides, the Residue-Similarity Index (RSI) can be computed as $\displaystyle \text{RSI} = 1 - |\text{RI} - \text{SI}|$. These scores or indices can be utilized to assess the performance of a given machine-learning features. Specially, in the current study, we use them to measure certain feature representation derived from scRNA-seq data. We name this type of analysis as RS analysis.

\subsection{Machine learning algorithms}

Machine learning modeling provides accurate, efficient and robust predictions. The original scRNA-seq data is not immediately suitable for machine learning. We integrate the proposed differential geometry curvature representation $\mathcal{C}$ of cell-cell interactions with machine learning algorithms, resulting in MDG predictive models. In this work, we mainly utilize gradient boosting decision tree (GBDT) as our machine learning algorithm.  The hyperparamters we employed in GBDT model are: $\rm n{\_}estimators=2000, max{\_}depth=7, min{\_}samples{\_}split=5, learning{\_}rate=0.002, subsample=0.8,  max{\_}features = "sqrt"$. In the following section, we will demonstrate the predictive performance of our machine learning models.

\subsection{Evaluation metrics }

In this study, we test the performance of our machine learning models mainly on some classification tasks for scRNA-seq data. We demonstrated the effectiveness our models by comparing to some state-of-the-art methods. Commonly used evaluation metrics in classification problems include recall, precision, F1-score, and AUC. The definitions for them are given below:
\begin{align*}
	&\text{Recall} = \text{True Positive}/(\text{True Positive}+ \text{False Negative})\\
	&\text{Precision} =\text{True Positive}/(\text{True Positive}+ \text{False Positive}) \\
	&\text{F1-Score}= 2\times ( \text{Precision} \times \text{Recall}) /( \text{Precision}+ \text{Recall})
\end{align*}
In the context of imbalanced datasets, where the number of instances in each class is not equal, macro-metrics can be useful. Single-cell RNA sequencing (scRNA-seq) data often exhibits class imbalance, where certain cell types may be less frequent than others. Macro-metrics in the multi-class classification are used to calculate performance metrics on a per-class basis and then average them, providing an overall performance measure. Macro-recall, macro-precision, macro-F1, and macro-AUC are often considered metrics. Assuming that there are $c$ classes in a given dataset, we have definitions for the several macro metrics as below:
\begin{align*}
	&\text{macro-Recall} = \frac{1}{c}\times \sum_{i=1}^{c}\text{Recall}_i,\quad
	\text{macro-Precision} = \frac{1}{c}\times \sum_{i=1}^{c}\text{Precision}_i\\
&\text{macro-F1}= \frac{1}{c}\times \sum_{i=1}^{c}\text{F1}_i,\quad
 \text{macro-AUC}=  \frac{1}{c}\times \sum_{i=1}^{c}\text{AUC}_i
\end{align*}
In the analysis of single-cell RNA sequencing data, the macro metrics provide a more balanced assessment of a model's effectiveness across different classes, ensuring that the performance on minority classes is not overshadowed by the majority ones.

\section{Results}\label{sec:results}
\subsection{Single-cell RNA sequencing datasets}

We can utilize the above differential geometry curvatures to analyze single cell RNA sequencing data (scRNA-seq). Due to the high dimensionality of scRNA-seq data, dimensionality reduction techniques, such as principal component analysis (PCA), t-distributed stochastic neighbor embedding (t-SNE), non-negative matrix factorization (NMF), and UMAP can be used to project original data into low-dimensional space, giving us low-dimensional embeddings.

We benchmark our multiscale models against these dimensionality reduction techniques on some scRNA-seq datasets to show the representative power of MDG. We build machine learning models by integrating these low-dimensional embeddings with GBDT algorithm. Twelve scRNA-seq datasets are used in the benchmark test, and their details can be found in Table \ref{tab: dataset}. The data was normalized using either reads per kilobase of transcript per million (RPKM), transcript per million (TPM) or counts per million (CPM). Notably, the cell type column indicate the number of cell types in each dataset and they served as the classification labels in our machine learning modeling. 

\begin{table}[H]
	\centering
	\caption{Accession ID, source organism, and the counts for samples, genes, cell types and normalization for 12 datasets}
	\begin{tabular}{c|c|c c c c  c c |} \hline
		&Accession ID & Reference & Organism & Samples & Genes & Cell types  & Normalization \\ \hline
		1&GSE45719  & Deng \cite{deng2014single}& Mouse & 300 & 22431 & 8 & RPKM\\
		2&GSE67835 & Darmanis \cite{darmanis2015survey} & Human & 420 & 22084 & 8 & CPM\\
		3&GSE75748 cell & Chu \cite{chu2016single} & Human & 1018 & 19097 & 7 & TPM\\
		4&GSE75748 time & Chu \cite{chu2016single} & Human & 758 & 19189 & 6 & TPM \\
		5&GSE82187 & Gokce \cite{gokce2016cellular} & Mouse & 705 & 18840 & 10 & TPM\\
		6&GSE84133 h1 & Baron \cite{baron2016single} & Human & 1937 & 20125 & 14 & TPM\\
		7&GSE84133 h2 & Baron\cite{baron2016single} & Human & 1724 & 20125 & 14 & TPM\\
		8&GSE84133 h3 & Baron\cite{baron2016single} & Human & 3605 & 20125 & 14 & TPM\\
		9&GSE84133 h4 & Baron\cite{baron2016single} & Human & 1308 & 20125 & 14 & TPM\\
		10&GSE84133 m1 & Baron\cite{baron2016single} & Mouse & 822 & 14878 & 13 & TPM\\
		11&GSE84133 m2 & Baron\cite{baron2016single} & Mouse & 1064 & 14878 & 13 & TPM\\
		%12&GSE89232 & Breton \cite{breton2016human} & Human & 957 & 20689 & 4 & TPM \\
		12&GSE94820 & Villani \cite{villani2017single} & Human & 1140 & 26593 & 5 & TPM\\ \hline
	\end{tabular}
	\label{tab: dataset}
\end{table}

\begin{figure}[h]
	\centering
	\includegraphics[width=0.9\linewidth]{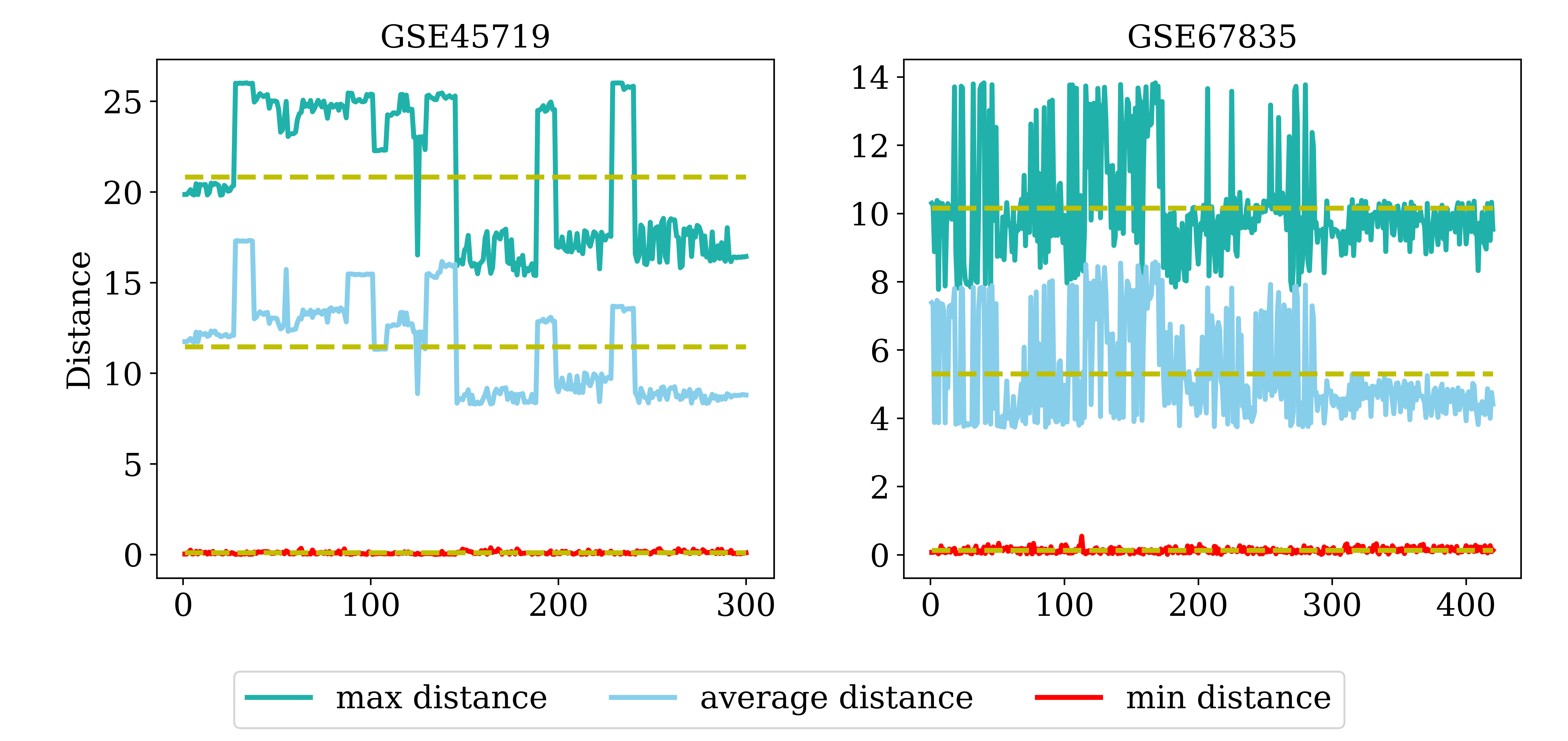} 
	\caption{{\footnotesize The statistical analysis of sample distances within dataset GSE45719 and GSE67835 based on the CCP-UMAP representations. The resulting plots display the maximum, average, and minimum distances of each sample from other cells, represented by light blue, dark blue, and red curves, respectively. Additionally, a yellow line is included to denote the average values for each of these three distance metrics. The $x$-axis indicates the row index of cells in the original scRNA-seq file. }}
	\label{Fig:distance-analysis-example}
\end{figure}

\subsection{MDG and benchmark machine learning models}

Classifying cell types is an important task in the study of scRNA-seq data. Machine learning models based on our MDG features can serve as useful tools in this regard. We demonstrate the predictive ability of MDG through classification predictions.

In our  MDG model, we select a variety of $\eta$ and $\kappa$ values. We restrict two bounds for $\eta$ value as in condition \eqref{eq:etabounds}. It is critical to analyze the CCP-UMAP representation $X$ to determine $\eta$ bounds. Figure \ref{Fig:distance-analysis-example} illustrates the distributions of minimum, average, and maximum distances between the samples and other cells in dataset GSE45719. The range of distances falls between 0.095 and 20.83. We consider a maximum value of 21 for the parameter $\eta$ in the exponential kernel. For $\eta$, we utilize a range of values from 1 to 21 in increments of 0.5, i.e., $\eta = 1,1.5,2,\cdots,21$. The parameter $\kappa$ is set at 5 and 10. Similarly, for dataset GSE67835 with minimum and maximum distances of 0.13 and 10.16, the minimum  and maximum $\eta$ values are set to 1 and 10, respectively. Additional distance distributions for the remaining datasets are presented in Figure S1. As all 12 datasets exhibit a minimum distance value close to 0, we set a common minimum $\eta$ value of 1. The strategy for setting $\eta$ and $\kappa$ values remains consistent across these datasets. 

Our CCP-UMAP representation $X$ offer a three dimensional data for differential geometry modeling. To make fair comparisons, we use dimensionality reduction algorithms to project original scRNA-seq data to three dimensions. For the aforementioned four dimensionality reduction techniques, the original data was log-transformed, and genes with variance less than $10^{-6}$ were removed prior to the reduction. We name the resulting features as PCA, UMAP, tSNE, and NMF, corresponding to the dimensionality reduction algorithms.

To asses our machine learning models, we test five-fold cross validations using a variety of evaluation metrics as mentioned before. To mitigate the effects of randomness, we construct machine learning models ten times with different random seeds and then compare the averages of the evaluation scores.

\subsection{MDG aids in cell type classification with scRNA-seq data}

\begin{figure}[h]
	\centering
	\includegraphics[width=0.92\linewidth]{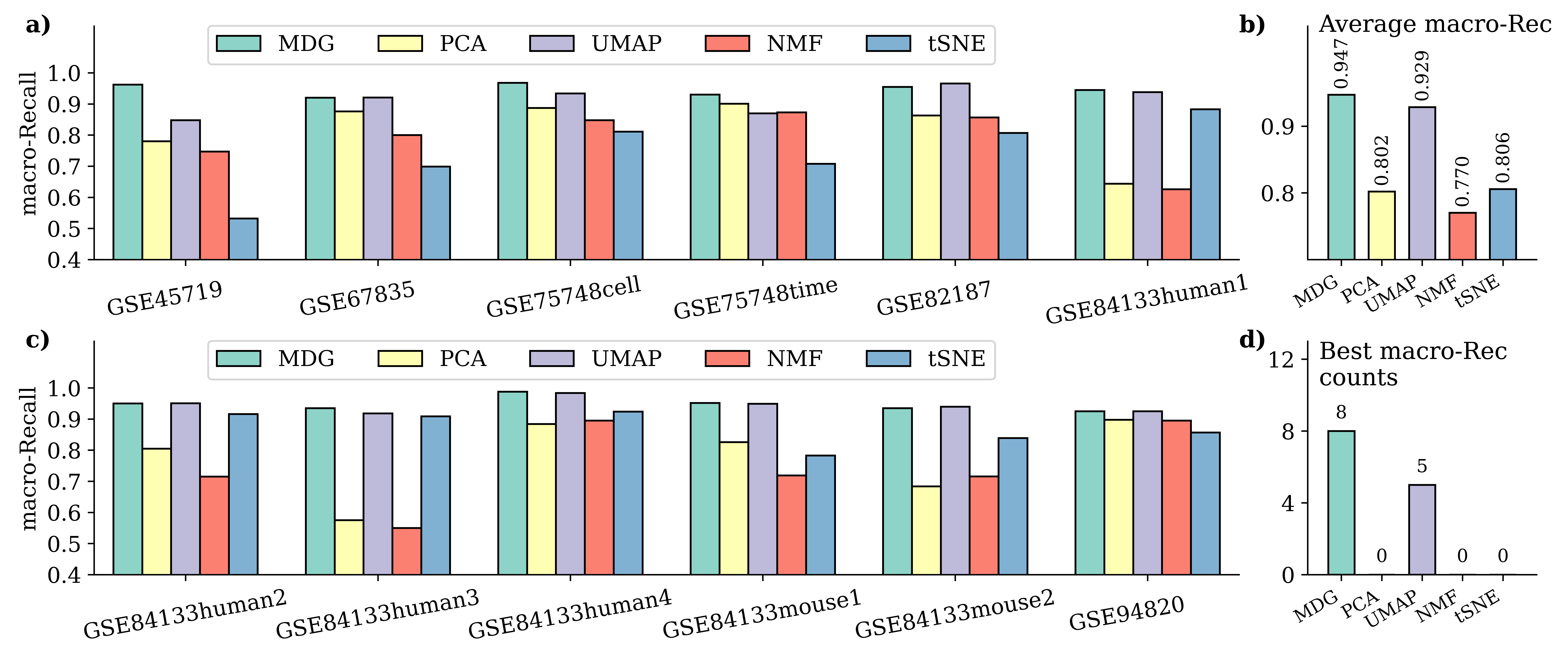} 
	\caption{{\footnotesize The macro-Recall comparisons of our MDG model with other models for 12 cell type classification tasks. Other models are using dimensionality reduction embeddings from PCA, UMAP, NMF, and tSNE. Our MDG model achieves the highest macro-Recall of 0.947 and got the best results for 8 out of the 12 tasks. }}
	\label{fig:RECALL-comparision-DGMM-reductions}
\end{figure}

In this section, we demonstrate the performance of MDG  models in cell type classifications. Table \ref{tab:RECALL-comparision-DGMM-reductions} shows the comparisons between our models with other models. It can be observed that our MDG model achieves the highest average macro-Recall value of 0.9472 for the 12 datasets. The macro-Recall score is identical to balanced accuracy in multi-classification prediction problems, which is the average of the true positive rate across all classes. The MDG model gives macro-Recall values of over 0.9 for all the 12 datasets, which indicates the model robustness for the cell type classification tasks. It achieves the best predictions for 11 of the 12 datasets compared to the other four models in terms of macro-Recall as shown in Figure \ref{fig:RECALL-comparision-DGMM-reductions} or Table \ref{tab:RECALL-comparision-DGMM-reductions}. The average macro-Recall score by MDG model is $1.2\%$, $18.1\%$, $23.0\%$, and $17.6\%$ higher than UMAP, PCA, NMF, and tSNE models. 

Out of the four dimensionality reduction methods, UMAP  model has the highest  average macro-Recall score of 0.9211. It is not as effective as our MDG model, but is still far superior to the other three dimensionality reduction methods. The comparisons in terms of macro-Recall, macro-AUC, and macro-F1 show that UMAP  model achieves the best predictions for 5, 5, and 5 tasks among the 12 prediction tasks. MDG model win most of the remaining tasks except that the tSNE model give a best prediction based on macro-AUC. UMAP is recognized for its ability to preserve both local and global structures when projecting high-dimensional data into a lower-dimensional space \cite{mcinnes2018umap}. Its performance with current embedding confirms its dimensionality reduction power. However, UMAP has relatively poor performance for GSE45719 and GSE75748time datasets with macro-Recall values below 0.9. It is likely due to that the two datasets have large amounts of gene yet small numbers of samples.

tSNE  models give the second best predictions among the four dimensionality reduction methods. It has an average macro-Recall of 0.8057, which is much lower than those by MDG and UMAP  models. The model based on NMF  data performs the worst among these models. The average macro-Recall value for these classification tasks is only 0.7701. This indicates that NMF  data is not suitable in analyzing scRNA-seq data. 

%tSNE can only project high-dimensional data to either two or three dimensional space. In our test, models with two dimensional tSNE embedding have better performance than 3D counterparts, as seen in Table \ref{tab:RECALL-comparision-DGMM-reductions} and \ref{tab:BA-comparision-DGMM-ccp-3D}. It may render more reliable 2D visualization than its 3D data. 
 
 The PCA models has poorer performance than MDG, UMAP, and tSNE. They can give good performance for dataset GSE84133human2 and GSE94820. But the average macro-Recall value for these dataset is 0.8019. PCA is a linear dimensionality reduction technique. It is effective at capturing linear relationships and structures in the data, while the scRNA-seq data that is both complex and non-linear. The current PCA-reduction data is not well suitable to characterize those non-linear interactions between cells. Despite that tremendous popularity of PCA in single cell data analysis, its low dimension reduction is not suitable for machine learning analysis for the single cell data through our investigations.

\begin{figure}[ht]
	\centering
	\includegraphics[width=0.92\linewidth]{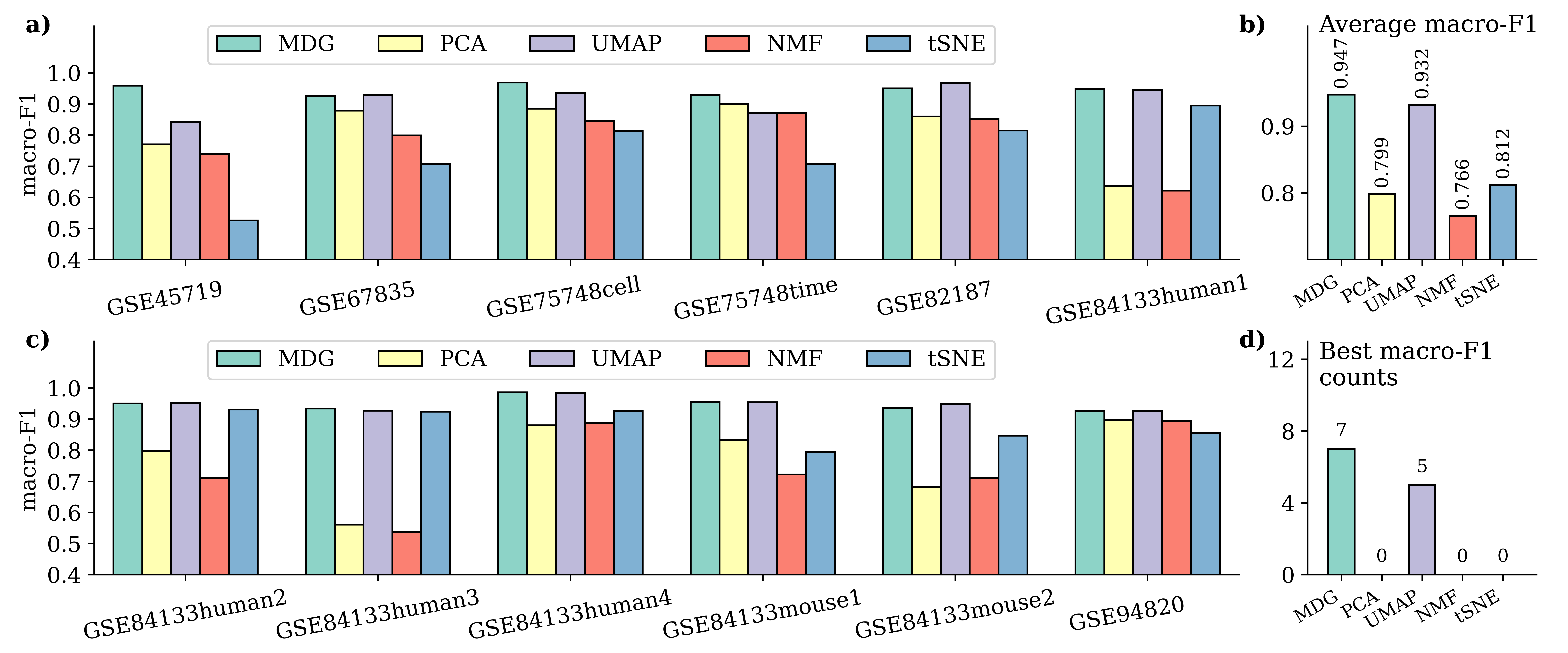} 
	\caption{{\footnotesize The Macro-F1 comparisons of our MDG model with other models for 12 cell type classification tasks. Other models are using dimensionality reduction embeddings from PCA, UMAP, NMF, and tSNE. Our MDG model achieves the highest F1 of 0.947 and got the best results for 7 out the 12 tasks.}}
	\label{fig:F1-comparision-DGMM-reductions}
\end{figure}

\begin{figure}[h]
	\centering
	\includegraphics[width=0.92\linewidth]{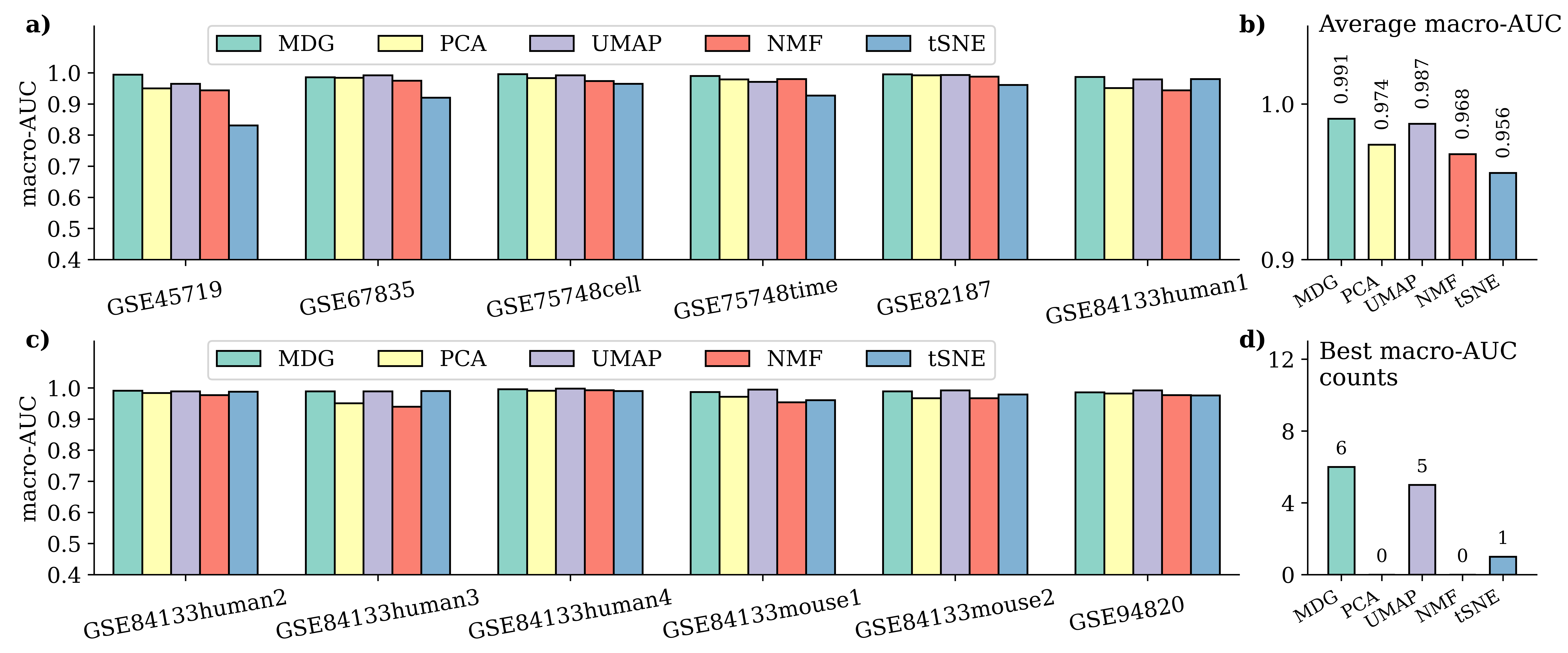} 
	\caption{{\footnotesize The Macro-AUC comparisons of our MDG model with other models for 12 cell type classification tasks. Other models are using dimensionality reduction embeddings from PCA, UMAP, NMF, and tSNE. Our MDG model achieves the highest AUC of 0.991 and got the best results for 6 out of the 12 tasks. }}
	\label{fig:AUC-comparision-DGMM-reductions}
\end{figure}

The comparisons between MDG model with other models in terms of macro-F1 and macro-AUC are presented in Figures \ref{fig:F1-comparision-DGMM-reductions} and \ref{fig:AUC-comparision-DGMM-reductions} or Tables \ref{tab:F1-comparision-DGMM-reductions} and \ref{tab:AUC-comparision-DGMM-reductions}. DG-MM attains the top 7 macro-F1 scores among the 12 tasks and the top 6 macro-AUC scores across the 12 tasks. These findings further highlight the advantages of MDG models over others. Additional comparisons of these model with accuracy and macro-Precision are provided in section S2 of the Supporting information.

 The CCP-UMAP features are used in our MDG modeling. These features can be directly employed in machine learning modeling. We evaluate performance of its models against those using dimensionality reduction embeddings, and table S3 presents the results of these comparisons. Additionally, we  comparisons of the MDG model with these models are illustrated in Tables S3 and S4. All these comparisons demonstrated the outstanding performance of the MDG modeling.

	\begin{table}[h]
	\centering
	\begin{tabular}{c|c|c|c|c|c|c|}
		\hline
		
		&	\multicolumn{1}{c|}{\multirow{2}{*}{\bf{Datasets}}} 	&\multicolumn{5}{c|}{Macro-Recall comparison}\\
		\cline{3-7}
		& \multicolumn{1}{c|}{ }& MDG & PCA  &  UMAP  & NMF &  tSNE  \\
		\hline		
1&GSE45719      &\textbf{0.962}&0.78&0.848&0.747&0.532\\
2&GSE67835      &0.92&0.876&\textbf{0.921}&0.8&0.699\\
3&GSE75748cell  &\textbf{0.968}&0.887&0.934&0.848&0.811\\
4&GSE75748time  &\textbf{0.93}&0.901&0.87&0.873&0.708\\
5&GSE82187      &0.955&0.863&\textbf{0.966}&0.857&0.807\\
6&GSE84133human1&\textbf{0.945}&0.644&0.938&0.626&0.883\\
7&GSE84133human2&0.95&0.805&\textbf{0.951}&0.715&0.916\\
8&GSE84133human3&\textbf{0.935}&0.575&0.918&0.55&0.909\\
9&GSE84133human4&\textbf{0.988}&0.884&0.984&0.895&0.924\\
10&GSE84133mouse1&\textbf{0.952}&0.826&0.949&0.719&0.783\\
11&GSE84133mouse2&0.935&0.684&\textbf{0.94}&0.716&0.839\\
12&GSE94820      &\textbf{0.925}&0.898&\textbf{0.925}&0.895&0.857\\
\hline
&Average macro-Recall&\textbf{0.9471}&0.8019&0.9287&0.7701&0.8057\\
		\hline		
	\end{tabular}
	\caption{The macro-Recall comparisons of our MDG model with other models in cell type classification predictions. Other models are using features from dimensionality reduction algorithms including PCA, UMAP, NMF, and tSNE. }
	\label{tab:RECALL-comparision-DGMM-reductions}
\end{table}

\begin{table}[h]
	\centering
	\begin{tabular}{c|c|c|c|c|c|c|}
		\hline
		
		&	\multicolumn{1}{c|}{\multirow{2}{*}{\bf{Datasets}}} 	&\multicolumn{5}{c|}{Macro-F1 comparison}\\
		\cline{3-7}
		& \multicolumn{1}{c|}{ }&MDG&PCA &UMAP &NMF &tSNE \\
		\hline		
		1&GSE45719      &\textbf{0.959}&0.77&0.842&0.739&0.526\\
		2&GSE67835      &0.926&0.879&\textbf{0.929}&0.799&0.707\\
		3&GSE75748cell  &\textbf{0.969}&0.885&0.936&0.846&0.814\\
		4&GSE75748time  &\textbf{0.929}&0.901&0.871&0.872&0.708\\
		5&GSE82187      &0.95&0.86&\textbf{0.968}&0.852&0.815\\
		6&GSE84133human1&\textbf{0.949}&0.636&0.946&0.622&0.895\\
		7&GSE84133human2&0.95&0.798&\textbf{0.952}&0.71&0.931\\
		8&GSE84133human3&\textbf{0.934}&0.561&0.927&0.538&0.924\\
		9&GSE84133human4&\textbf{0.986}&0.88&0.984&0.888&0.926\\
		10&GSE84133mouse1&\textbf{0.955}&0.834&0.954&0.722&0.794\\
		11&GSE84133mouse2&0.936&0.682&\textbf{0.948}&0.71&0.847\\
		12&GSE94820      &0.925&0.896&\textbf{0.926}&0.893&0.855\\
		\hline
		&Average Macro-F1&\textbf{0.9473}&0.7985&0.9319&0.7659&0.8118\\
		\hline		
	\end{tabular}
	\caption{The Macro-F1 comparisons of our DGMM model with other models in cell type classification predictions. Other models are using features from dimensionality reduction algorithms including PCA, UMAP, NMF, and tSNE.}
	\label{tab:F1-comparision-DGMM-reductions}
\end{table}

	\begin{table}[h]
	\centering
	\begin{tabular}{c|c|c|c|c|c|c|}
		\hline
		
		&	\multicolumn{1}{c|}{\multirow{2}{*}{\bf{Datasets}}} 	&\multicolumn{5}{c|}{Macro-AUC comparison}\\
		\cline{3-7}
		& \multicolumn{1}{c|}{ }&MDG&PCA &UMAP &NMF &tSNE \\
		\hline		
1&GSE45719      &\textbf{0.994}&0.95&0.965&0.944&0.831\\
2&GSE67835      &0.986&0.984&\textbf{0.992}&0.975&0.92\\
3&GSE75748cell  &\textbf{0.996}&0.983&0.992&0.974&0.965\\
4&GSE75748time  &\textbf{0.99}&0.979&0.971&0.98&0.927\\
5&GSE82187      &\textbf{0.995}&0.992&0.993&0.988&0.961\\
6&GSE84133human1&\textbf{0.987}&0.951&0.979&0.944&0.98\\
7&GSE84133human2&\textbf{0.991}&0.984&0.989&0.977&0.988\\
8&GSE84133human3&0.989&0.951&0.989&0.94&\textbf{0.99}\\
9&GSE84133human4&0.996&0.991&\textbf{0.998}&0.993&0.99\\
10&GSE84133mouse1&0.987&0.972&\textbf{0.995}&0.954&0.961\\
11&GSE84133mouse2&0.989&0.967&\textbf{0.992}&0.967&0.979\\
12&GSE94820      &0.986&0.982&\textbf{0.992}&0.977&0.976\\
\hline
&Average Macro-AUC&\textbf{0.9905}&0.9738&0.9872&0.9678&0.9557\\
		\hline		
	\end{tabular}
	\caption{The Macro-AUC comparisons of our DGMM model with other models in cell type classification predictions. Other models are using features from dimensionality reduction algorithms including PCA, UMAP, NMF, and tSNE.}
	\label{tab:AUC-comparision-DGMM-reductions}
\end{table}

\subsection{RS analysis} 

In our previous studies, we found that RSI is associated with classification accuracy \cite{yuta2022ccp,yuta2023singlecell}.  RSI uses the features and labels to compute the scores. In addition, residue-similarity (R-S) plot can serve as visualization tool for the scRNA-seq data. We use it to visualize the predictions by our models for each class. The residue (R) and similarity (S) scores are computed for each sample based on the given features for the scRNA-seq data.

\begin{figure}[ht]
	\centering
	\includegraphics[width=0.88\linewidth]{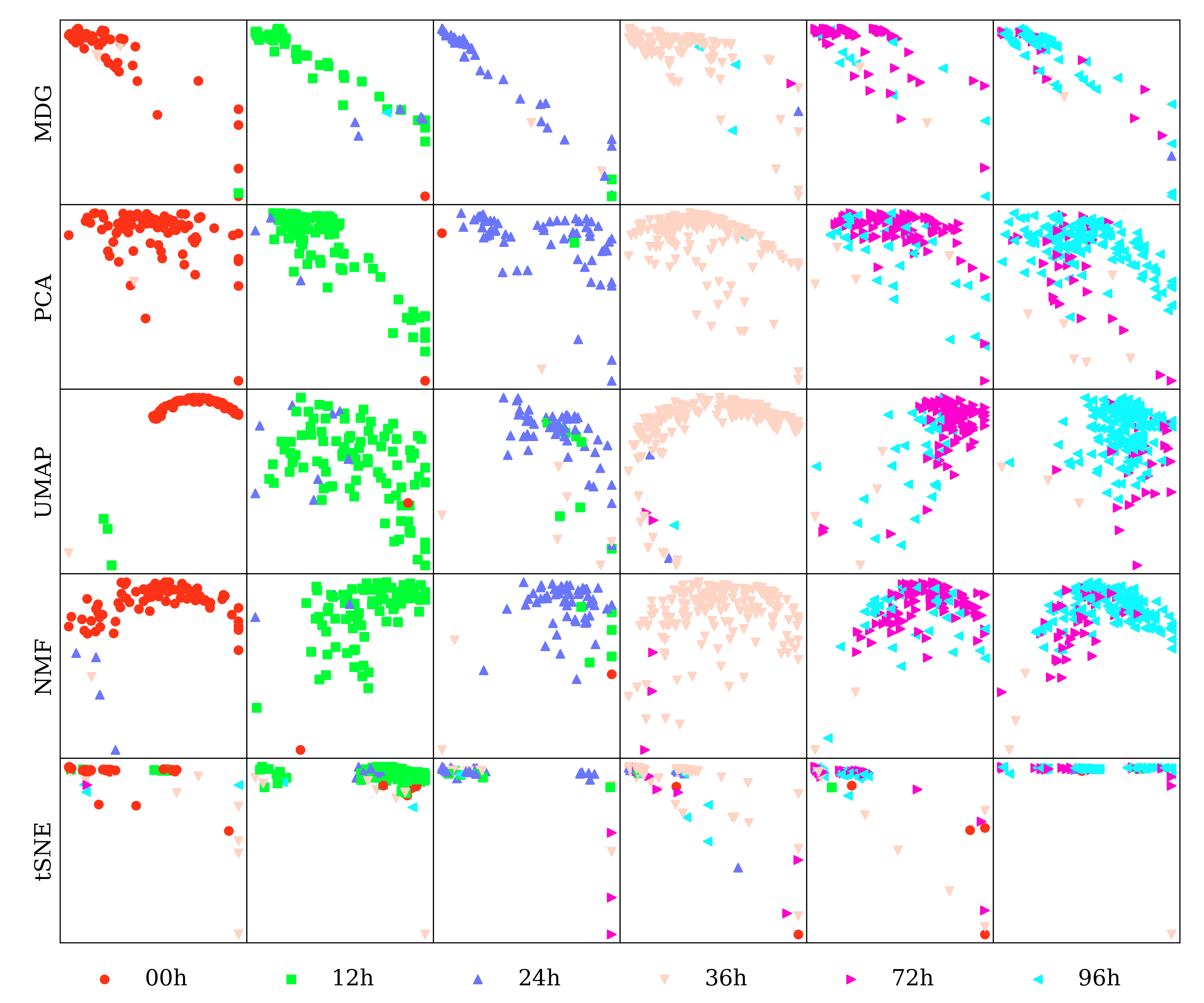} 
	\caption{{\footnotesize The rs-plot of MDG, PCA, UMAP, NMF, and tSNE features for GSE75748 time data. The columns indicate different classes of cell types in the original scRNA-data, while the rows correspond to the five models. }}
	\label{Fig:Rs-plot-GSE75748time-dggl-reduction}
\end{figure}

 In our study, we compared MDG modeling with four other dimensionality reduction models. We assess the five kinds of embeddings via RS analysis. We used five-fold cross-validation to divide the feature data into five parts, where four parts were used to train GBDT models, and one part was used as a test set. Then, R and S scores were computed for each sample in the five test sets. In the RS-plot, the x-axis and y-axis correspond to R and S scores of the samples, respectively. In the plots, samples were colored according to their predicted labels from the GBDT classifier. Figure \ref{Fig:Rs-plot-GSE75748time-dggl-reduction} shows a comparison between the R-S plot of MDG, PCA, UMAP, NMF, and tSNE embeddings for GSE75748 time data. The columns indicate different classes of cell types in the original scRNA-data, while the rows correspond to the five models.

 Both residue and similarity scores range from 0 to 1, where 1 is the most optimal. If the samples cluster in the top-right corner, it indicates that samples within a class closely resemble each other, and different classes are clearly distinct from one another. Consequently, RSI can be correlated with classification accuracy. The RSI for MDG, PCA, UMAP, NMF, and tSNE  models are 0.454, 0.820, 0.978, 0.932, and 0.584. Despite the low RSI of our DG-MM features, its integration with GBDT still offers superior predictions. This is attributed to the merits of GBDT algorithm in reducing overfitting. This reflects the descriptive ability of DG-MM features for cell-cell interactions despite that it suffers from overfitting issues. UMAP features gives the highest RSI, which is consistent with its superior classification predictions over than dimensionality reduction modeling. 

Figure \ref{Fig:Rs-plot-GSE75748time-dggl-reduction} shows the prediction results for GSE75748 time data by the five models. The macro-Recall scores are 0.93, 0.901, 0.97, 0.873, and 0.708 for MDG, PCA, UMAP, NMF, and tSNE models, respectively. These models all show poor predictions for classifying 72h and 96h cell types. According to \cite{chu2016single}, GSE75748 time data records ES cell differentiation from pluripotency to definitive endoderm (ED) at 0th hour, 12th hour, 24th hour, 36th hour, 72th hour, and 9th hour over 4 days. The misclassification of the 72th hour, and 96th hour data is consistent with the finding that cells from 72th hour and 96th hour were relatively homogeneous \cite{chu2016single}. Our MDG model provide better predictions for the cell types at the two times than the other four models as seen in the last two columns in Figure \ref{Fig:Rs-plot-GSE75748time-dggl-reduction}. There are various misclassifications as seen in the R-S plots. Through the visualization, we can observed the relatively better classification by our model while the tSNE model performs the worst.

\section{Discussion}

\begin{figure}[ht]
	\centering
	\includegraphics[width=0.85\linewidth]{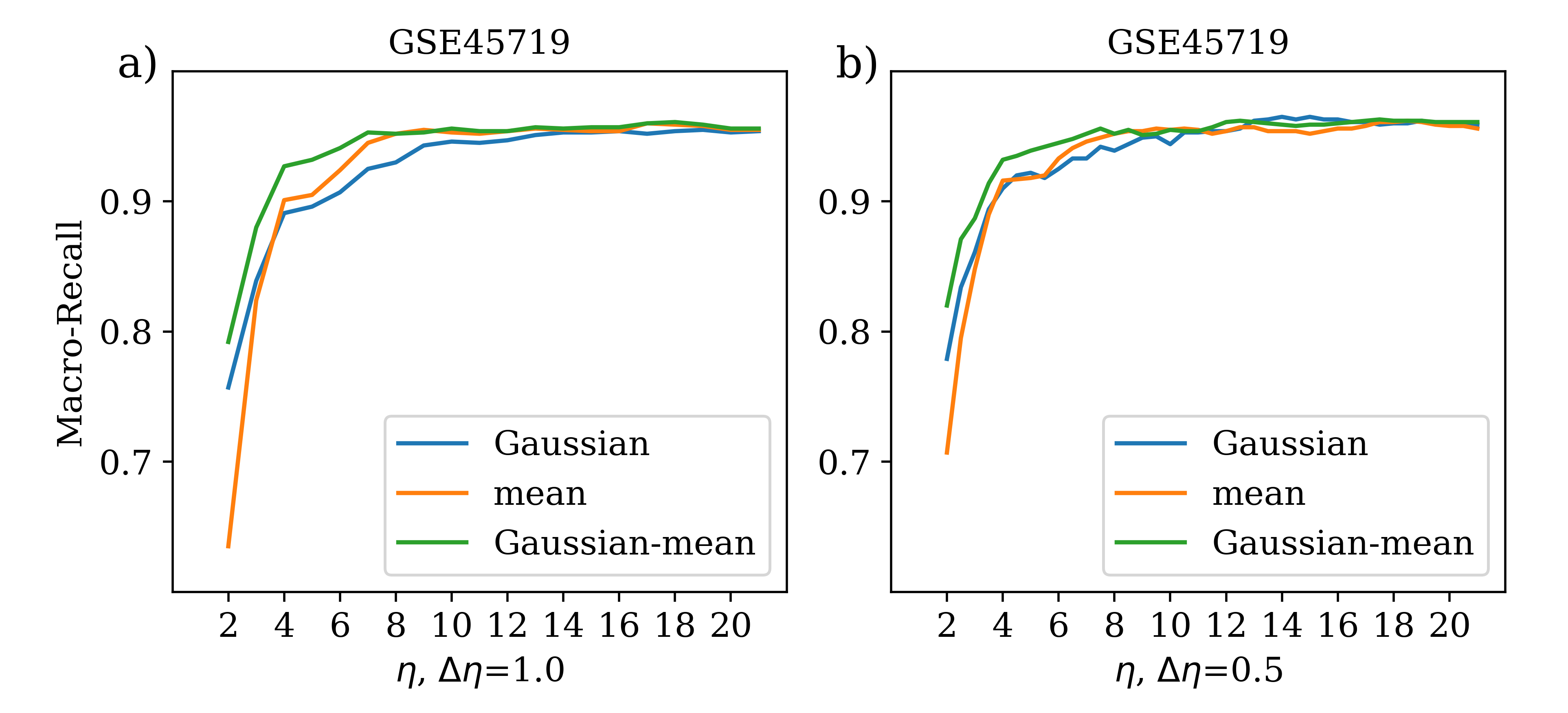} 
	\caption{{\footnotesize The multiscale modeling performance is evaluated based on accumulated curvatures. Gaussian curvature, mean curvature, or their concatenations are the three types of features utilized. The parameter dx, set equal to either 1 or 0.5, determines the step size for collecting curvature at a $\eta$, and incorporating it into the final features. }}
	\label{Fig:multiscale-comparisions}
\end{figure}

Our   multiscale differential geometry modeling is accomplished by collecting Gaussian and mean curvatures at various $\eta$ values. The chosen $\eta$ values span a range between 1 and the mean maximum pairwise distance in the CCP-assisted UMAP embeddings. The selected step size $\Delta \eta$  is 0.5. Additionally, we assess the performance by employing a subset of these resultant features in various aspects. Figure \ref{Fig:multiscale-comparisions} displays the macro-Recall scores of five-fold cross validation classifications for dataset GSE45719 using various embeddings.

In addition to concatenating Gaussian and mean curvatures, we assessed performance using either Gaussian or mean curvatures individually. For the GSE45719 dataset, we initially computed both types of curvatures at $\eta$ values ranging from 1 to 21, with a step size $\Delta \eta$ equal to either 1 or 0.5, i.e. $\eta_i=1,1+\Delta \eta,\cdots,21$. Correspondingly, we obtain Gaussian and mean curvatures $(K^{\eta_i},H^{\eta_i})$. We collected the curvatures with $\eta$ starting at 1 and concluding at a predetermined value greater than 1 and less than 21. As illustrated in Figure \ref{Fig:multiscale-comparisions}, the utilization of additional features results in improved performance across all three scenarios, whether employing Gaussian curvature, mean curvature, or their concatenations. On the other hand, the concatenation of Gaussian and mean curvatures shows improved performance compared to using only Gaussian or mean curvatures, especially when a smaller concluding value, $\eta_i$, is employed. However, as the concluding value $\eta_i$ increases, the concatenation does not give significant advantages over solely using Gaussian or mean curvatures. 

These multiscale curvature-based embedding reach their predictive power plateau around $\eta_i=12$ for all the three scenarios. As shown in Figure \ref{Fig:distance-analysis-example}, the mean of average pairwise distance in CCP-assisted UMAP embedding is around 12. This suggest that concluding $\eta$ value can be set to the mean of average distance rather than mean of maximal distance in Eqn \ref{eq:etabounds}. The curvature features based on the $\eta$ value greater than 12 in the current case can be redundant, but the GBDT algorithm have metrics of prioritizing important features and reducing overfitting, which make our curvature embedding robust in the classification predictions. In practice, it is sufficient to choose either Gaussian or mean curvatures with the concluding $\eta$ value set to be the mean of average distance in the CCP-assisted UMAP embeddings. But the addition of more Gaussian and mean concatenation curvatures cause slightly improved predictions. 

Figures \ref{Fig:multiscale-comparisions}a and \ref{Fig:multiscale-comparisions}b compare predictions made with accumulated curvature features using different values of $\Delta \eta$. Smaller values of $\Delta \eta$ tend to result in better predictions. In the instance of concatenating Gaussian and mean curvatures, the macro-Recall is 0.962 with $\Delta \eta=0.5$, whereas the macro-Recall is 0.956 with $\Delta \eta=1$. Consequently, the $\Delta \eta$ value of 0.5 is a preferred option in our multiscale modeling.

\section{Conclusion}

In this study, we developed a multiscale differential geometry (MDG) model for analyzing and interpreting scRNA-seq data. We employed correlated clustering and projection (CCP)-assisted UMAP (CCP-UMAP) methodology to preprocess scRNA-seq data. CCP is our recently proposed dimensionality reduction method that projects each cluster of similar genes into a super-gene defined as accumulated pairwise nonlinear gene-gene correlations among cells. Its integration with UMAP projects the original scRNA-seq data onto a 3D space that faithfully preserve enough geometric and biological information inherent in scRNA-seq data. We assume that those intrinsic properties lies on a family of low-dimensional manifolds embedded in the high-dimensional scRNA-seq data. The multiscale low-dimensional manifolds are constructed by discrete-to-continuum mapping based on the derived CCP-UMAP embeddings and are then used to capture the interactive cell-cell interactions in the cell network. Gaussian and mean curvatures from differential geometry offer quantitative estimates of the cell-cell interactions and give rise to multiscale differential geometry embeddings for scRNA-seq data. The cell-cell interactions within these manifolds preserve the neighboring similarity of the original data, revealing the complex topological structure among cells. This, in turn, allows for the elucidation of cell–cell relationships in the network. 

 We compared its representative ability with embeddings from standard dimensionality reduction methods in classifying cell types. These embeddings are paired with gradient boosting decision tree clustering algorithm to provide machine learning predictions. Their comparison demonstrate the utilization of multiscale curvatures can be an effective tool of scRNA-seq data interpretation. A multiscale analysis was provided to showcase the necessity of various cell-cell interactive manifolds utilized, which was reflected by the enhanced clustering performance resulting from the accumulation of additional curvatures. 

The multiscale differential geometry analysis  serves as an effective approach to decipher distinct cell features and cell–cell relationships from the high dimensional space of scRNA-seq data. The complex topological and geometric structures are well captured. The scRNA-seq data is characterized by high-dimensional data at single cell level for a population of cells, which gives a foundation for dissection of cell heterogeneity. Many other challenging problems exist in disciplines such image analysis, environmental science, social science, astronomy and astrophysics involving point specific high-dimensional data. Dimensionality reduction or networks analysis tools are needed to address these challenges. Our MDG approach is not restricted to the analysis of scRNA-seq data; it also holds the potential to be applied to a variety of other networks in science and engineering.

\section{Code and Data availability}
All data and the code needed to reproduce this paper's result can be found at\\ \hyperlink{https://github.com/WeilabMSU/MDG}{https://github.com/WeilabMSU/MDG}. \\
CCP is made available through our web-server at  \\ \hyperlink{https://weilab.math.msu.edu/CCP/}{https://weilab.math.msu.edu/CCP/} or through the source code \hyperlink{https://github.com/hozumiyu/CCP}{https://github.com/hozumiyu/CCP}. \\
Source code of RSI and R-S plot can be found at \hyperlink{https://github.com/hozumiyu/RSI}{https://github.com/hozumiyu/RSI}. 

\section{Acknowledgments}
This work was supported in part by NIH grants R01GM126189, R01AI164266, and R35GM148196, NSF grants DMS-2052983, DMS-1761320, DMS-2245903,  and IIS-1900473, NASA grant 80NSSC21M0023, MSU Foundation, Bristol-Myers Squibb 65109, and Pfizer.

%\section*{References}
%\bibliographystyle{unsrt}
%\bibliographystyle{plain}
%\bibliography{references}

\end{document}